\documentclass[floats,floatfix,amssymb,prd,twocolumn,superscriptaddress,nofootinbib]{revtex4-2}

\usepackage{amssymb,amsmath,verbatim,mathtools,needspace,enumitem,etoolbox,graphicx,physics,microtype,afterpage,bm}
\usepackage[dvipsnames, usenames]{xcolor}
\definecolor{linkcolor}{rgb}{0.0,0.3,0.5}
\usepackage[unicode, colorlinks=true, linkcolor=linkcolor, citecolor=linkcolor, filecolor=linkcolor,urlcolor=linkcolor, pdfusetitle]{hyperref}
\usepackage[all]{hypcap}
\usepackage[T1]{fontenc}
\usepackage[utf8]{inputenc}
\usepackage{tabularx}
\usepackage{float}
\usepackage{lmodern}

\allowdisplaybreaks

\usepackage{tikz}
\usepackage{color}
\usepackage{framed}
\usepackage{hyperref}
\hypersetup{colorlinks, citecolor=bluscuro, linkcolor=black, urlcolor=bluscuro}

\usepackage{bm}
\usepackage{latexsym}
\usepackage{amsmath}
\usepackage{amssymb}
\usepackage{amsthm}
\usepackage{amscd}
\usepackage[mathscr]{eucal}
\usepackage{graphicx}
\usepackage{subfigure}
\usepackage{psfrag}
\usepackage{rotating}

\usepackage{color}

\definecolor{ggreen}{cmyk}{0.7,     0,      0.9,      0}
\definecolor{viol}{cmyk}{0.3,1,0,0}
\definecolor{myred}{cmyk}{0.1, 1, 0.5, 0}
\definecolor{bblue}{rgb}{0.2, 0.29996, 0.8 }

\definecolor{rossos}{cmyk}{0,1,1,0.55}
\definecolor{bluscuro}{rgb}{0.15, 0.2, .85}
\definecolor{bluchiaro}{cmyk}{1,.3,0.,0.1}

\definecolor{ForestGreen}{rgb}{0.13, 0.55, 0.13}

\theoremstyle{plain}

\begin{document}


\title{Scale invariant elastic stars in General Relativity}

\author{Artur Alho}
\affiliation{Center for Mathematical Analysis, Geometry and Dynamical Systems, Instituto Superior T\'ecnico, Universidade de Lisboa, Av. Rovisco Pais, 1049-001 Lisboa, Portugal}

\author{Jos\'e Nat\'ario}
\affiliation{Center for Mathematical Analysis, Geometry and Dynamical Systems, Instituto Superior T\'ecnico, Universidade de Lisboa, Av. Rovisco Pais, 1049-001 Lisboa, Portugal}

\author{Paolo Pani}
\affiliation{Dipartimento di Fisica, Sapienza Universit\`a di Roma \& INFN Roma1, Piazzale Aldo Moro 5, 00185, Roma, Italy}

\author{Guilherme Raposo}
\affiliation{Departamento de Matem\'atica da Universidade de Aveiro and Centre for Research and Development in Mathematics and Applications (CIDMA), Campus de Santiago, 3810-183 Aveiro, Portugal}


\begin{abstract}
We present a model of relativistic elastic stars featuring scale invariance. 
This implies a linear mass-radius relation and the absence of a maximum mass.
The most compact spherically symmetric configuration that is radially stable and satisfies all energy and causality conditions has a slightly smaller radius than the Schwarzschild light ring radius. To the best of our knowledge, this is the first material compact object with such remarkable properties in General Relativity, which makes it a unique candidate for a black-hole mimicker. 
\end{abstract}

\maketitle

\section{Introduction}
Within General Relativity~(GR), 
black holes~(BHs) are fascinating and quite unique solutions to the vacuum Einstein equations $R_{\mu\nu}=0$. Since the Ricci tensor $R_{\mu\nu}$ is scale-free, the BH mass $M$ is a free parameter of the solution. This implies that, at least in principle, BHs can form with any mass. 
In spherical symmetry, their horizon radius is proportional to their mass, $R=2M$ (we use units such that $G=c=1$), and so we have the (dimensionless) compactness ${\cal C}\equiv M/R=1/2$ for any BH.
Furthermore, there is strong evidence that GR BHs are stable~\cite{Regge:1957td,Whiting:1988vc,Dafermos:2016uzj,Klainerman:2017nrb,Hafner:2019kov,Dafermos:2021cbw,Klainerman:2021qzy,Giorgi:2022omp}.

There exists an active research program aiming to challenge the BH paradigm and test the nature of compact objects with the ever-growing wealth of astrophysical observations~\cite{Cardoso:2019rvt}.
One of the cornerstones of this program is developing models of BH mimickers (also known as exotic compact objects), namely horizonless, regular solutions to GR (or extensions thereof) that replace the curvature singularity unavoidably present inside BHs~\cite{HawkingEllis,1969NCimR...1..252P} with a regular matter content, while mimicking the peculiar BH properties as closely as possible.

However, coupling to matter almost inevitably introduces a scale into the Einstein equations, which implies that the mass of the solutions is not a free parameter anymore.
Well-known examples are white dwarfs and neutron stars, whose mass scale is set by the Chandrasekhar limit~\cite{Chandrasekhar:1931ih}, $M_{\rm Chandra}\sim \hbar^{3/2}/m_N^2\approx 1.4 M_\odot$, where $m_N$ is the nucleon mass.
More exotic examples are the self-gravitating geons known as boson stars~\cite{Liebling:2012fv}, whose mass scale is set by the Kaup limit~\cite{Kaup:1968zz}, $M_{\rm Kaup}\sim \hbar/m_B\approx( 10^{-10}\,{\rm eV}/m_B)M_\odot $, where $m_B$ is the boson mass (this bound is modified in case of strong bosonic self-interactions~\cite{Colpi:1986ye,Lee:1991ax}).
The presence of a mass scale is very general, and implies the existence of a maximum mass above which the solution is unstable against radial perturbations, and either migrates towards less massive, stable, configurations, or collapses into a BH.
Thus, an exotic compact object can at most mimic a BH in some mass range, but it cannot encompass the whole range $M\in(\sim 1 M_\odot,\sim 10^{10} M_\odot)$ where astrophysical BHs are expected and indeed observed.
For the same reason, in classical GR the merger of two horizonless compact objects, both of which are near the maximum mass, is bound to form a BH, providing a strong argument in favor of the existence of BHs even when very compact horizonless objects do exist in the universe.

The main goal of this work is to present the first GR model that challenges this paradigm. The key idea is to find a matter content allowing for \emph{scale invariance} within GR. This guarantees that the mass is a free parameter of the solution, and that the mass-radius relation is linear at any scale.
As we shall show, such model exists within the framework of relativistic elastic materials, and, furthermore, allows for solutions which are radially stable and physically admissible regarding causality limits and energy conditions.

\section{Elastic stars}

In this section we summarize the main features of the theory of relativistic elasticity. There are several formulations of relativistic elasticity in the literature, starting with the foundational work of Carter and Quintana~\cite{CarterQuintana}, and more recently by Kijowski and Magli~\cite{KijMag92} and Beig and Schmidt~\cite{Beig:2002pk}. Here we follow closely the formalism put forward by Beig and Schmidt and the novel formalism developed in our previous work~\cite{Alho:2021sli,Alho:2022bki} (see~\cite{Alho:2023ris} for a detailed description, \cite{Beig:2002pk,Beig:2023pka} for an introduction to relativistic elasticity theory, and \cite{Par00,Karlovini:2002fc,Karlovini:2003xi,Karlovini:2004gq,FK07,Andreasson:2014lka} for other works on elastic relativistic stars). 


\subsection{Mapping formalism: physical spacetime and reference space}
The configuration of a relativistic elastic body is described by a projection map $\bm{\Pi}:\mathcal{S}\to\mathcal{B}$, mapping the physical spacetime $(\mathcal{S},\bm{g})$, where the physical, deformed, object exists, to the 3-dimensional  Riemannian material space $(\mathcal{B},\bm{\gamma})$, where the object is in its undeformed reference state. The corresponding push-forward map $d\bm{\Pi}: \mathcal{T}\mathcal{S}\rightarrow \mathcal{T}\mathcal{B}$ is called the \emph{configuration gradient}.
We can assign local coordinates  $x^\mu$ to $\mathcal{S}$ and $X^I$ to $\mathcal{B}$, so that the mapping $\bm{\Pi}$ reads $X^I = \Pi^I(x^\mu)$, and the configuration gradient $d\bm{\Pi}$ is $f^I_\mu = \partial_\mu \Pi ^I$.  The inverse images by $\bm{\Pi}$ of the points in $\mathcal{B}$ are assumed to form a congruence of timelike curves in $\mathcal{S}$, generated by the $4$-velocity $u^\mu$ of the particles making up the body, so that
\begin{equation}\label{Orthof}
u^{\mu}\partial_\mu \Pi^I =0 \, .
\end{equation}  

In relativistic elasticity, the concept of elastic strain is intimately related to the push-forward of the inverse spacetime metric $H^{IJ} = f_\mu^I f_\nu^J g^{\mu\nu}$. It is convenient to define the Riemannian metric $h_{\mu\nu}$ induced on the subspaces orthogonal to the 4-velocity of the particles $u^\mu$,
\begin{equation}
    h_{\mu\nu} = f_\mu^I f_\nu^J H_{IJ} = g_{\mu\nu} + u_{\mu} u_{\nu} \, .
\end{equation}
%

\subsection{Lagrangian formalism and stress-energy tensor}

The dynamics of relativistic elastic bodies can be obtained from a variational principle for the action
\begin{equation}
S[\bm{\Pi}] =\int_{\mathcal{S}}\rho(\bm{\Pi},d\bm{\Pi})\sqrt{-\det(\bm{g})}\,d^4x \, ,
\end{equation}
where $\rho$ is the energy density of the matter.

It follows from this variational principle and from the diffeomorphism invariance that one can write the stress-energy tensor as
\begin{equation}
T_{\mu\nu}=\rho u_\mu u_\nu +\sigma_{\mu\nu} \, , \qquad  \sigma_{\mu\nu}u^{\nu}=0 \, ,
\end{equation}
where 
\begin{equation}\label{eq:Cauchy_initial}
\sigma_{\mu\nu}=2 \frac{\partial\rho}{\partial g^{\mu\nu}}-\rho h_{\mu\nu}
\end{equation}
is the symmetric Cauchy stress tensor.
The diffeomorphism invariance assumption also entails that the energy density should not depend on the configuration gradient explicitly, but instead one should have $\rho(\bm{\Pi},\bm{H})$. 

We further restrict our analysis to homogeneous and isotropic materials. For the latter, the energy density depends only on the push-forward metric via its principal invariants,
\begin{equation}
\rho=\rho (i_1(\bm{H}),i_2(\bm{H}),i_3(\bm{H})) \, ,
\end{equation}
which are given in terms of the eigenvalues $h_1, h_2, h_3$ of $H^I_J = H^{IK}\gamma_{KJ}$ by the usual symmetric polynomials $i_1 = h_1 + h_2 + h_3$, $i_2=h_1 h_3 + h_1 h_2 + h_2h_3$ and $i_3 = h_1h_2h_3$. These eigenvalues are positive, and can be seen as the squares of the normalized linear particle densities $n_1$, $n_2$, and $n_3$ along the principal directions spanned by the eigenvectors $e^{J}_{(i)}$, $i=1,2,3$,  since~\cite{Alho:2023ris}
\begin{equation}\label{PLD}
\frac{n}{n_0}=\sqrt{h_1 h_2 h_3}=n_1 n_2 n_3,
\end{equation}
where $n$ and $n_0$ are the the number density of particles in the physical and reference state, respectively.

For these materials it is possible to simplify the symmetric Cauchy stress tensor~\eqref{eq:Cauchy_initial} to obtain,
\begin{equation}
\sigma_{\mu\nu} = \sum^{3}_{i=1} \left(n_i \frac{\partial\rho}{\partial n_i}-\rho\right) e_{(i)\mu}e_{(i)\nu},
\end{equation}
where $e_{(i)\mu}:=f^{I}_{\mu}e_{(i)I}$  is the pull-back of the orthonormal coframe $e_{(i)J}$.
It is clear that the Cauchy stress tensor is diagonal in this frame, and so we can identify the principal pressures as
\begin{equation}\label{eq:pressuregeneral}
    p_i := \left(n_i \frac{\partial\rho}{\partial n_i}-\rho\right)\,.
\end{equation}

\subsection{The wave speeds}
For elastic materials there are exactly 9 independent wave speeds, corresponding to \emph{longitudinal} waves in $i$-th direction,
\begin{equation}
c^2_{\mathrm{L}i} =\frac{\displaystyle n_i \frac{\partial p_i}{\partial n_i}}{\rho+p_i},
\end{equation} 
and to \emph{transverse} waves in $i$-th direction, oscillating in the $j$-th direction:
\begin{equation}\label{TWaves}
c^2_{\mathrm{T}ij} = 
\begin{cases}
\displaystyle\frac{n^2_j }{\rho+p_j}\frac{(p_j-p_i)}{n^2_j-n^2_i}, & \text{if}\ n_i\neq n_j ,\\ \\
\displaystyle\frac{\frac{1}{2}n_j}{\rho+p_j}\left(\frac{\partial p_i}{\partial n_i}-\frac{\partial p_j}{\partial n_i}\right), & \text{if}\ n_i = n_j .
\end{cases}
\end{equation}

\subsection{Spherically symmetric compact elastic stars}
We now restrict the previous discussion to the case of spherical symmetry.
From the eigenvalues of $H^I_J$ we find that in spherical symmetry, $n_2=n_3$ (and $p_2=p_3)$. It is convenient to introduce the following change of variable: 
\begin{equation}\label{DefVars}
\varrho = \varrho_0 n_r n_t^2\, , \qquad\varsigma = \varrho_0 n_t^3 \, ,
\end{equation}
where we have defined $n_r=n_1$ and $n_t =n_2=n_3$. The quantity $\varrho$ is the baryonic mass density while the quantity $\varsigma$ is given in terms of the differential equation,
\begin{equation} \label{ODEvarsigma}
    \partial_r \varsigma = -\frac{3}{r}\left(\varsigma - \frac{\varrho}{(1-2m/r)^{1/2}}\right),
\end{equation}
which is obtained from the eigenvalues of $H^A_B$. By looking at the integral form of Eq.~\eqref{ODEvarsigma},
\begin{equation}\label{DefVarsigma}
    \varsigma(r)=\frac{3}{r^3}\int^{r}_{0}\frac{\varrho(u)u^2du}{\left(1-{2m(u)}/{u}\right)^{1/2}}\,,
 \end{equation}
 we can interpret $\varsigma$ as the average mass density within a sphere of radius $r$. 

The radial and tangential pressures can be defined as $p_\mathrm{rad}=p_1$ and $p_\mathrm{tan}=p_2=p_3$.
%
%
In the spherically symetric case, Eqs.~\eqref{eq:pressuregeneral} simplify to
\begin{subequations}
\begin{align}
 \label{eq:rho} 
    \widehat{p}_\mathrm{rad}(\varrho,\varsigma)&=\varrho \partial_\varrho \widehat{\rho}(\varrho,\varsigma) - \widehat{\rho}(\varrho,\varsigma) \,, \\
    \widehat{p}_\mathrm{tan}(\varrho,\varsigma) &= \widehat{p}_\mathrm{rad}(\varrho,\varsigma) + \frac32 \varsigma \partial_\varsigma \widehat{\rho}(\varrho,\varsigma)\,
\end{align}\label{eq:EoS}
\end{subequations}
(where $\hat\rho$, $\hat{p}_{\rm rad}$ and $\hat{p}_{\rm tan}$ are the quantities $\rho$, $p_{\rm rad}$ and $p_{\rm tan}$ written as functions of $\varrho$ and $\varsigma$). Thus, the $\varsigma$-dependence of $\widehat{\rho}$ is generically associated with anisotropic matter, and the perfect fluid limit is obtained in the case $\partial_\varsigma \widehat{\rho}=0$, where one retrieves the isotropic pressure $\widehat{ p}_\mathrm{iso}(\varrho)\equiv\widehat{p}_\mathrm{rad}(\varrho)\equiv\widehat{p}_\mathrm{tan}(\varrho)$.

We consider the stars to be in hydrostatic equilibrium, so that the spacetime metric can be written as ${(g_{\mu\nu})={\rm diag}(-e^{2\phi(r)}, (1-2m(r)/r)^{-1}, r^2,r^2\sin^2\theta)}$.
The matter sector is generically described by the stress-energy tensor ${(T^{\mu}_{\,\,\nu})={\rm diag}(\rho, p_\mathrm{rad}, p_\mathrm{tan},p_\mathrm{tan})}$, where $\rho$, $p_\mathrm{rad}$ and $p_\mathrm{tan}$ are radial functions. 
Einsteins' field equations imply 
\begin{subequations}\label{TOVeq}
\begin{align}
& \frac{dm}{dr}=4\pi r^2 \rho\, , \\
& \frac{d\phi}{dr}= \frac{{m}+{4\pi  r^3}p_\mathrm{rad}}{r(r-2m)}\, , \\
& \frac{dp_\mathrm{rad}}{dr} =\frac{2}{r}(p_\mathrm{tan}-p_\mathrm{rad})-(p_\mathrm{rad}+\rho)\frac{d\phi}{dr}\,.  
\end{align}
\end{subequations}
This system of equations is closed by prescribing an equation of state~(EoS). In our framework, this is done by providing the energy density functional $\hat\rho(\varrho,\varsigma)$, the parametric relations for the pressures~\eqref{eq:EoS} and the equation~\eqref{DefVarsigma} for $\varsigma$.
%

%
Once a given EoS is prescribed, the equations for the stellar structure are solved by imposing regularity of the metric and matter functions at the center of symmetry, which implies {${(\varrho(0),\varsigma(0))=(\varrho_c,\varrho_c)}$}. The radius $R$ of the star is then defined by $p_\mathrm{rad}(R)=0$.
Due to spherical symmetry, in the vacuum region $r>R$ the solution is the standard Schwarzschild metric.

In spherical symmetry there are five independent wave speeds, related to the speeds of longitudinal and transverse perturbations along the radial and tangential directions~\cite{Karlovini:2002fc}. 
Three of these quantities,  the speed of longitudinal waves in the radial direction $c_{\rm L}$, the speed of transverse waves in the radial direction $c_{\rm T}$ and the speed of transverse waves in the tangential direction and oscillating in the radial direction $\tilde{c}_{\rm T}$ can be defined from Eqs.~\eqref{eq:EoS} and their $\varrho$ and $\varsigma$ derivatives,
\begin{subequations}\label{soundspeeds1}
\begin{align}
c^2_\mathrm{L}(\varrho,\varsigma)&= \frac{\varrho\partial_\varrho \widehat{p}_{\mathrm{rad}}(\varrho,\varsigma)}{\widehat{\rho}(\varrho,\varsigma)+\widehat{p}_{\mathrm{rad}}(\varrho,\varsigma)}\,,\\
c^2_\mathrm{T}(\varrho,\varsigma)&= \frac{\widehat{p}_{\mathrm{tan}}(\varrho,\varsigma) - \widehat{p}_{\mathrm{rad}}(\varrho,\varsigma)}{\left(\widehat{\rho}(\varrho,\varsigma)+\widehat{p}_{\mathrm{tan}}(\varrho,\varsigma)\right)\left(1 - \left(\frac{\varrho}{\varsigma}\right)^2\right)}\,,\\
 \tilde{c}^2_\mathrm{T}(\varrho,\varsigma)&= \left(\frac{\varrho}{\varsigma}\right)^2\frac{\widehat{p}_{\mathrm{tan}}(\varrho,\varsigma)-\widehat{p}_{\mathrm{rad}}(\varrho,\varsigma)}{\left(\widehat{\rho}(\varrho,\varsigma)+\widehat{p}_{\mathrm{rad}}(\varrho,\varsigma)\right)\left(1 - \left(\frac{\varrho}{\varsigma}\right)^2\right)}\,,
\end{align}
\end{subequations}
while the other two independent velocities, the speed of longitudinal waves in the tangential direction $\tilde{c}_{\rm T}$, and the speed of transverse waves in the tangential direction oscillating in the tangential direction $\tilde{c}_{\rm TT}$, must be obtained from the general form of the energy density function, before reducing to spherical symmetry. 
If one starts from the reduced form, then the residual freedom allows us to freely choose one of these speeds as part of the model prescription, the other being fixed by this choice; in this case, the simplest option is the so-called \emph{natural choice}, introduced in~\cite{Alho:2021sli,Alho:2023ris}, which we will use in this work:
\begin{subequations}\label{cNC}
\begin{align}
    \tilde{c}_\mathrm{L}^2(\varrho,\varsigma) &= \frac{\varrho \partial_\varrho \widehat{p}_{\rm tan}(\varrho,\varsigma) + 3 \varsigma \partial_\varsigma \widehat{p}_{\rm tan}(\varrho,\varsigma) }{\widehat\rho(\varrho,\varsigma) + \widehat{p}_{\rm tan}(\varrho,\varsigma)}, \\
    \tilde{c}_\mathrm{TT}^2(\varrho,\varsigma) &= \frac{\frac32 \varsigma \partial_\varsigma \widehat{p}_{\rm tan}(\varrho,\varsigma) }{\widehat\rho(\varrho,\varsigma) + \widehat{p}_{\rm tan}(\varrho,\varsigma)}.
\end{align}
\end{subequations}
%

\section{Scale-invariant elastic materials}

A spherically symmetric matter source is said to be \emph{scale-invariant} if, under the scaling transformation given by $r\to A^{-1}\tilde{r}$,
with $A$ an arbitrary positive real number, the energy density, the radial and the tangential pressures transform as
\begin{equation}
    (\rho(r),p_\mathrm{rad}(r),p_\mathrm{tan}(r)) \to A^{2} (\tilde{\rho}(\tilde{r}),\tilde{p}_\mathrm{rad}(\tilde{r}),\tilde{p}_\mathrm{tan}(\tilde{r})). \label{rescal}
\end{equation}
It then follows that $m(r)\to A^{-1} \tilde{m}(\tilde{r})$ and $\phi(r)\to \tilde{\phi}(\tilde{r})$, so that the system~\eqref{TOVeq} is invariant under the scaling transformation.

For spherically symmetric perfect fluids, it is well known~\cite{TC1971} that scale invariance is achieved for a subclass of models with a linear EoS,
\begin{equation}\label{EoSFluid}
    \widehat{\rho}(\varrho)=\mathrm{n}\widehat{p}_\mathrm{iso}(\varrho) = \mathrm{n}\mathcal{K}\varrho^{1+\frac{1}{\mathrm{n}}},
\end{equation}
where $\mathcal{K}$ is a positive constant and $\mathrm{n}>0$ is the polytropic index. Such an EoS has constant sound speed 
$c^2_\mathrm{s}=dp_\mathrm{iso}/d\rho=1/\mathrm{n}$.
However, it does not support self-gravitating objects with finite radius~\cite{Collins:1985}.

A natural extension of Eq.~\eqref{EoSFluid} to the elastic setting consists in requiring that the isotropic pressure, defined as {${p_{\rm iso} = 1/3 (p_{\rm rad} + 2 p_{\rm tan})}$}, satisfies
\begin{equation}
\widehat{\rho}(\varrho,\varsigma) = {\rm n} \widehat{p}_{\rm iso}(\varrho,\varsigma) = \frac{\rm n}3 \left( \widehat{p}_{\rm rad}(\varrho,\varsigma) + 2\widehat{p}_{\rm tan} (\varrho,\varsigma)\right)\,. \label{rhoISO}
\end{equation}
Replacing $p_{\rm rad}$ and $p_{\rm tan}$ by Eqs.~\eqref{eq:EoS}, we can solve the resulting partial differential equation for $\rho$ to obtain the expression for the scale-invariant EoS in the elastic setting:
\begin{equation}\label{eq:scaleinvEOS}
    \widehat{\rho}(\varrho,\varsigma)= \mathrm{n}\mathcal{K}\varsigma^{1+\frac1{\rm n}} h(y) \,,
\end{equation}
where $y=\varrho/\varsigma$ and $h(y)$ is a free function that in the perfect fluid case~\eqref{EoSFluid} reduces to  $h(y)=y^{1+1/\mathrm{n}}$. 
It turns out that $\widehat{\rho}(\varrho,\varsigma)$ is a positive homogeneous function of degree $1+1/\mathrm{n}$, as in the perfect fluid case~\eqref{EoSFluid}.
In Appendix~\ref{app:EoSgeneral} we show that the class of EoS~\eqref{eq:scaleinvEOS} arises generically when imposing scale invariance.

A specific scale-invariant EoS corresponds to choosing the function $h(y)$. The most simple ansatz, that is also motivated by the fluid expression, is a power-law type function,
\begin{equation}
h(y) = \alpha_0 + \alpha_1 y + \alpha_2 y^{\beta_2} \,.
\end{equation}
The unknown constants $\alpha_i$ can be determined by the compatibility with
the isotropic state ($y=1$) and with the linear elasticity conditions~\cite{Alho:2021sli,Alho:2023ris},
\begin{equation}
h(1)=1, \quad h^\prime(1)=1+\frac{1}{\mathrm{n}}, \quad h^{\prime\prime}(1)=\frac{3}{\mathrm{n}}\left(1+\frac{1}{\mathrm{n}}\right)\left(\frac{1-\nu}{1+\nu}\right),  \nonumber 
\end{equation}
where $\nu\in(-1,1/2]$ is the Poisson ratio. Finally, we arrive at the form for the scale-invariant EoS:
\begin{widetext}
\begin{equation}\label{eq:prestreconstiso}
\widehat{\rho}(\varrho,\varsigma) =  \mathrm{n}{\cal K}\varsigma^{1+\frac{1}{\mathrm{n}}}\left[1 
	-\frac{1+\mathrm{n}}{\mathrm{n}}\left(1-3\frac{\mathrm{s}}{\mathrm{n}}\left(\frac{1-\nu}{1+\nu}\right)\right)\left(1-\frac{\varrho}{\varsigma}\right)-\frac{3(1+\mathrm{n})\mathrm{s}^2}{(1+\mathrm{s})\mathrm{n}^2}\left(\frac{1-\nu}{1+\nu}\right)\left(1-\left(\frac{\varrho}{\varsigma}\right)^{1+\frac{1}{\mathrm{s}}}\right)\right]\,,
\end{equation}
\end{widetext}
where we introduced the shear index $\mathrm{s}$ through $\beta_2 = 1 + \frac1{\rm s}$. 
The perfect fluid limit is obtained by taking $\mathrm{s}=\mathrm{n}$ and $\nu=1/2$, corresponding to the linear EoS~\eqref{EoSFluid}.

By computing the central density $\rho_c=\widehat{\rho}(\varrho_c,\varrho_c)$, central pressure $p_c =\widehat{p}_\mathrm{rad}(\varrho_c,\varrho_c)=\widehat{p}_\mathrm{tan}(\varrho_c,\varrho_c)$, and the wave speeds at center, we obtain inequalities giving necessary conditions for the energy and causality conditions to be satisfied:
\begin{equation}
	{\cal K}>0, \qquad\mathrm{n}\geq 3\left(\frac{1-\nu}{1+\nu}\right),\qquad -1<\nu\leq\frac{1}{2}\,.
   \end{equation}
%
%

 
\section{Numerical results}
%
\begin{figure*}[ht!]
	\centering
	\includegraphics[width=0.49\textwidth]{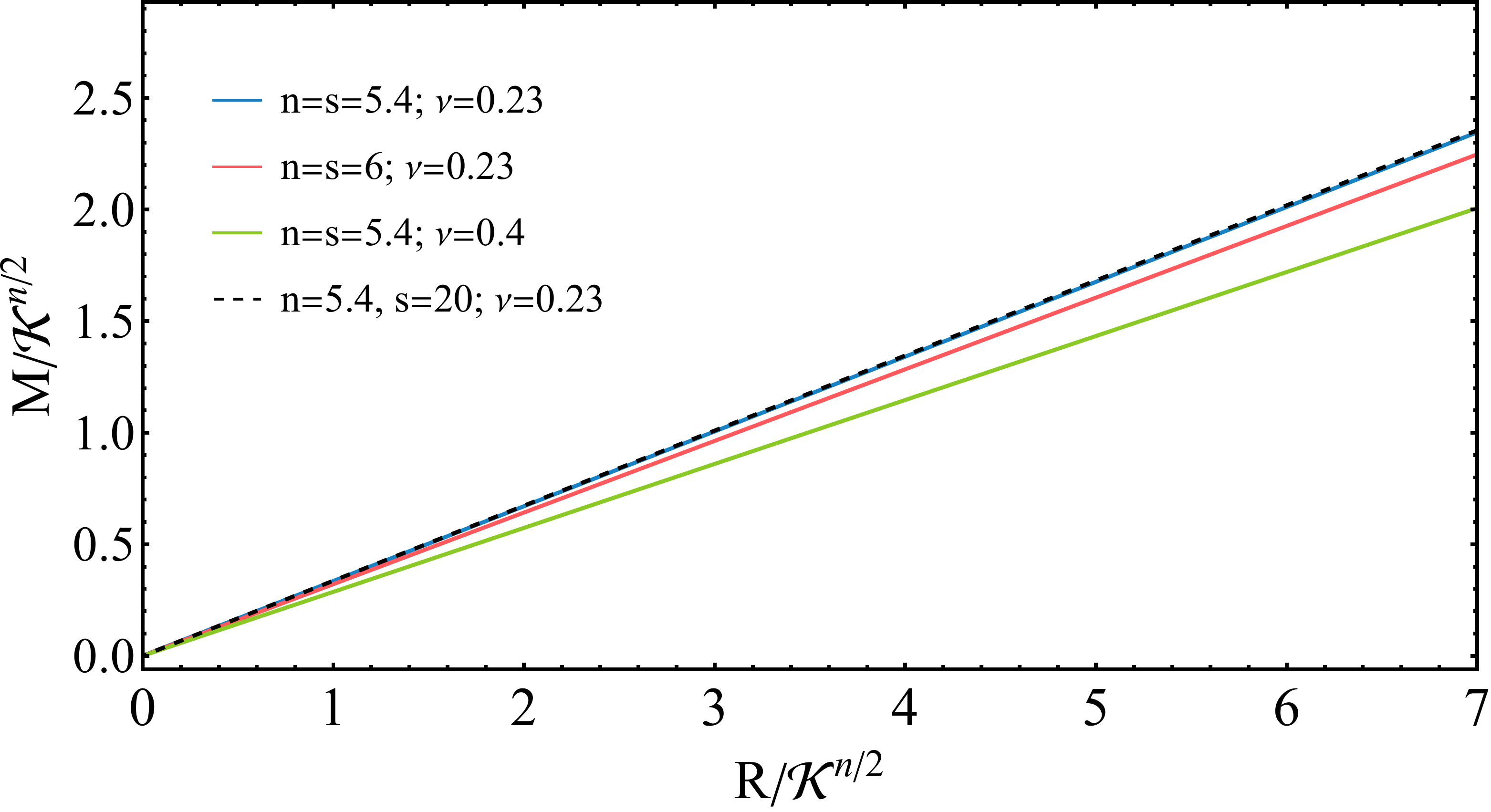} 
     \includegraphics[width=0.49\textwidth]{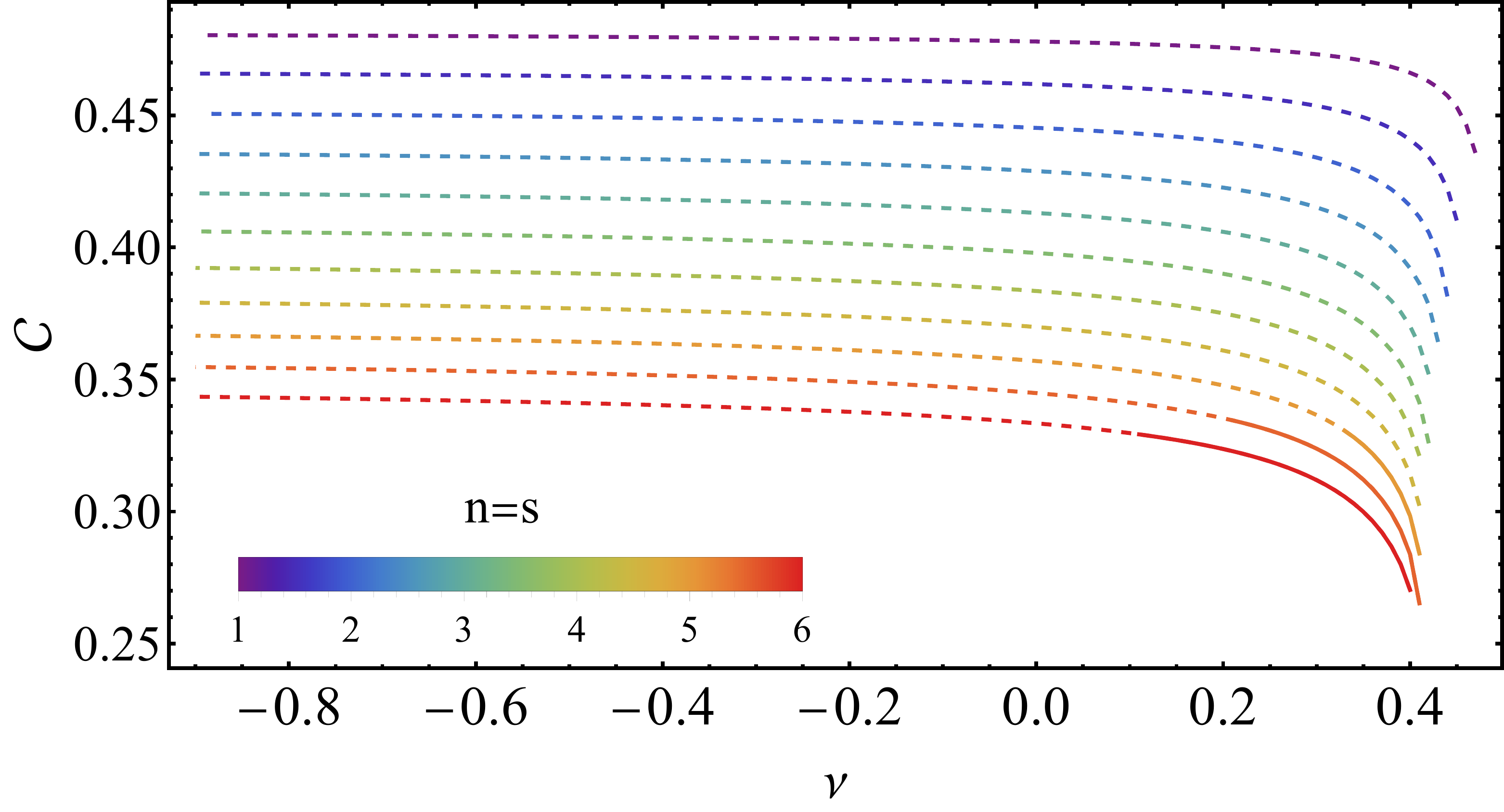} 
	\caption{ Left: Mass-radius diagram of our scale-invariant elastic material for representative values of the EoS parameters (n, s, $\nu$). Right: Compactness of different solutions as a function of the Poisson ratio $\nu$ and for different ${\rm n}={\rm s}$. Unphysical configurations (violating either the energy conditions or the causality bounds) are denoted by a dashed style.
	\label{fig:MRlinear}}
\end{figure*}
%
Self-gravitating perfect fluid solutions with a linear EoS and finite radius do not exist within GR~\cite{Collins:1985}. Remarkably, however, such bounded configurations do exist in this model for sufficiently high values of the elasticity. We show some representative examples in the mass-radius diagram on the left panel of Fig.~\ref{fig:MRlinear}. As anticipated above, due to the scale invariance of the model, the mass-radius relation is linear, and so the compactness is independent of the central value of the baryonic density: it only depends on the model parameters $(\mathrm{n},\mathrm{s},\nu)$, while the sole dimensionful parameter ${\cal K}$ simply sets a reference scale, as in the standard perfect fluid case in Newtonian gravity (see e.g.~\cite{ShapiroTeukolsky}). The compactness of representative solutions in this model is shown in the right panel of Fig.~\ref{fig:MRlinear} as a function of $\nu$ and for different values of ${\rm n=s}$. 
Note that, for fixed values of ${\rm n}$ and ${\rm s}$, elastic stars can only exist for $\nu\leq \nu_{\rm crit}({\rm n},{\rm s})<1/2$, in agreement with the fact that a certain amount of elasticity is required for their existence.

We observe that decreasing the Poisson ratio (i.e., increasing elasticity), or decreasing the index ${\rm n}$, increases the compactness of the solutions. Although the compactness  can reach very high values (up to the BH limit, ${\cal C}\to1/2$, for ${\rm n}\to0$), this happens for unphysical configurations where some of the sound speeds in the material exceed the speed of light (dashed branches). We find that physically admissible configurations (solid branches) require sufficiently high values of ${\rm n}$ and $\nu$. 

\begin{figure}[ht!]
	\centering
	\includegraphics[width=0.48\textwidth]{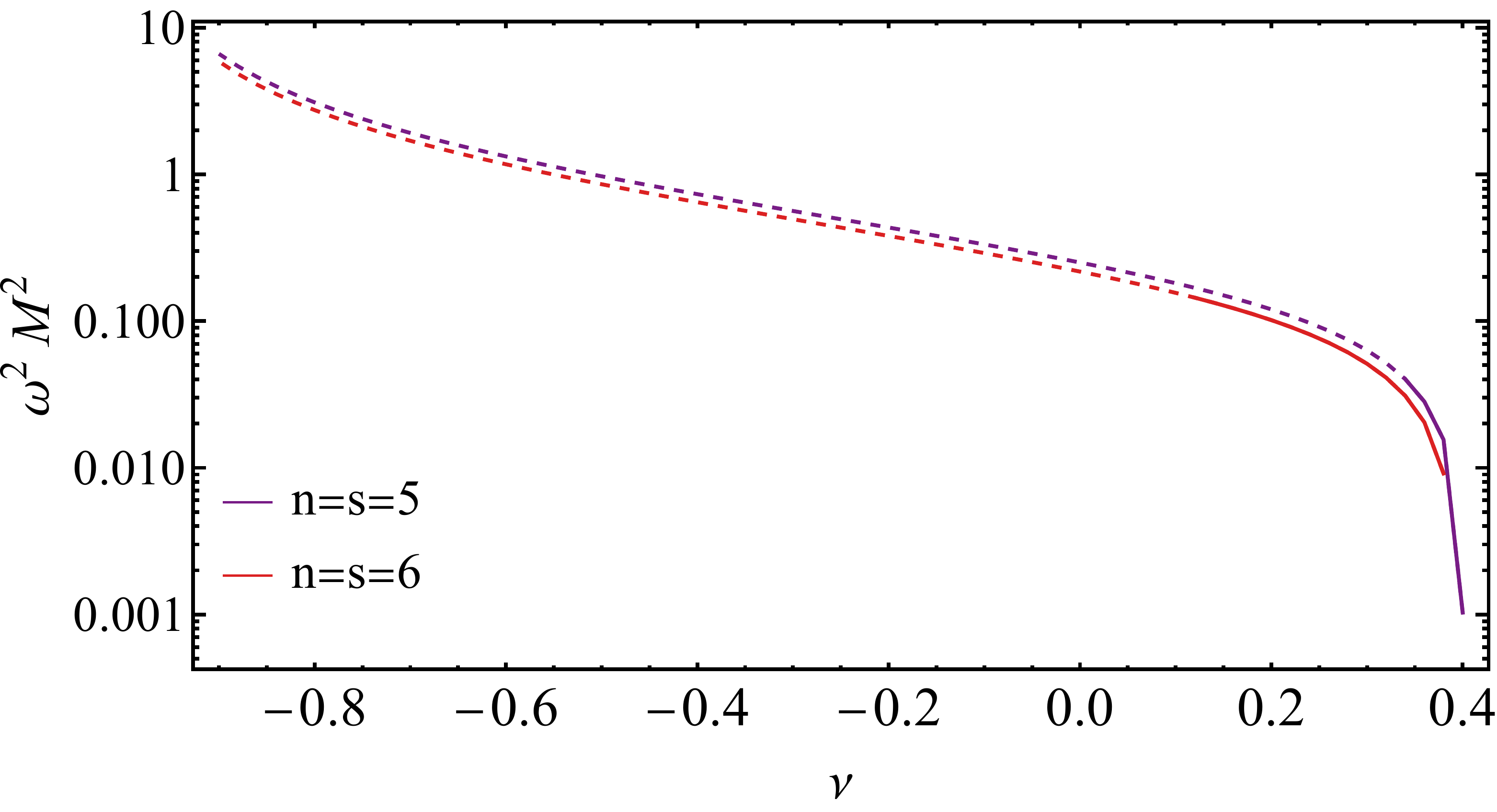} 
	\caption{Squared frequency of the fundamental mode for linear radial perturbations of scale-invariant elastic stars as a function of the Poisson ratio  $\nu$, for some representative values of ${\rm n}={\rm s}$. The dashed and solid lines correspond to unphysical and physically admissible configurations, respectively. In all cases, we found no evidence of unstable modes. \label{fig:stabilitylinear}
	}
\end{figure}

The linear mass-radius diagram is a truly remarkable feature, which, to the best of our knowledge, has never been reported for viable matter within GR. Scale invariance implies that there is no maximum mass in the model, so that these solutions can exist with any mass. These features make them akin to ordinary BHs, for which the mass is indeed a free parameter, and $M=R/2$ in the nonspinning case.

Prompted by these quite unique properties, we analyze the radial stability of these solutions using the (Eulerian) perturbation formalism developed in~\cite{Alho:2023ris}, which is based on 
introducing radial metric and matter perturbations with $\sim e^{-i
\omega t}$ time dependence and recasting the linearized Einstein equations into an eigenvalue problem for $\omega^2$~\cite{Karlovini:2003xi,Alho:2021sli}.
In Fig.~\ref{fig:stabilitylinear} we show the squared frequency of the fundamental mode, $\omega^2$, as a function of the Poisson ratio of the material for some representative values of (${\rm n},{\rm s}$). In all cases that we have numerically explored, we always found $\omega^2>0$, indicating that the solutions are radially stable. 

\begin{figure}[ht!]
	\centering
	\includegraphics[width=0.48\textwidth]{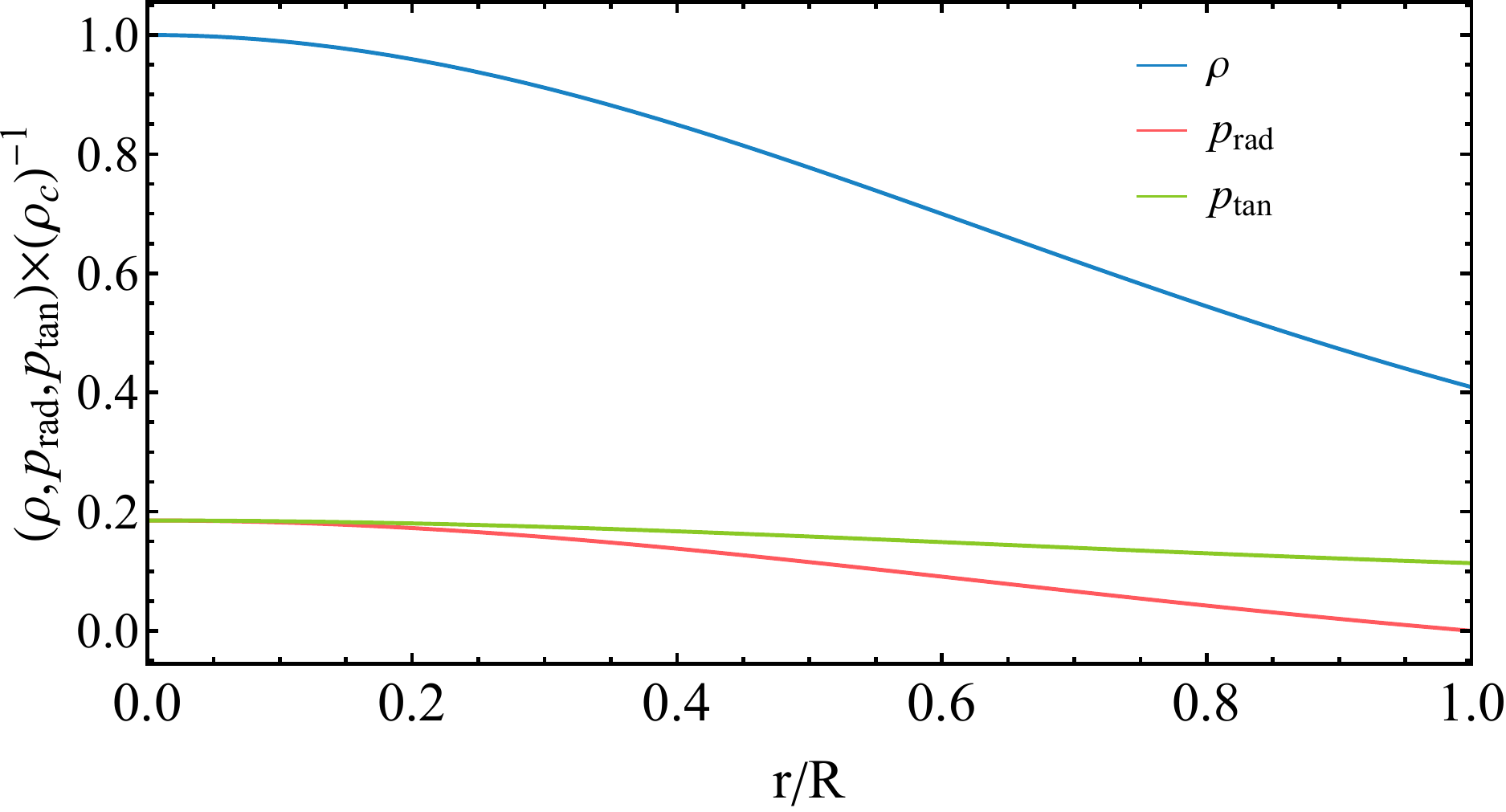} 
	\caption{Energy density, radial pressure, and tangential pressure profiles for the most compact physically admissible solution, corresponding to the compactness~\eqref{eq:boundlinearEOS}. By virtue of scale-invariance the normalized profiles do not depend on the value of central density.\label{fig:matterplots}
	}
\end{figure}

Therefore, the physically admissibility of the solution depends only on the subluminality of the wave speeds and 
on the energy conditions of the elastic material. 
By imposing these constraints, we obtain the maximum compactness of physically admissible configurations for this model,
\begin{equation}\label{eq:boundlinearEOS}
    {\cal C}_{\rm max}^{\rm PAS}\approx 0.335\,,
\end{equation}
which is saturated for $\rm n=s\approx 5.4$ and $\nu \approx 0.23$ (solid blue curve in the left panel in Fig.~\ref{fig:MRlinear}).
Intriguingly, the radius of the most compact physically admissible model is only slightly smaller than the light ring at $r=3M$. 
In Fig.~\ref{fig:matterplots} we show the profiles for the energy density, radial pressure, and tangential pressure for the configuration saturating the ${\cal C}_{\rm PAS}$ bound~\eqref{eq:boundlinearEOS}. In addition to satisfying causality and radial stability the solution also satisfies all energy conditions.

Finally, higher values of the shear index ${\rm s}$ tend to slightly increase the mass and the compactness of the solution, but also contribute to make the material superluminal. 
In the mass-radius diagram of Fig.~\ref{fig:MRlinear}, we compare the curve corresponding to the solution that approaches our bound~\eqref{eq:boundlinearEOS} (blue line) with a solution with the same parameters but a much larger shear index ${\rm s}$ (black dashed line). Although the mass-radius diagram is very similar in both cases, the solution with higher shear index is always superluminal near the radius of the star. While there is no \emph{a priori} upper limit for the shear index ${\rm s}$, its lower bound is fixed by the condition that the shear ratio on the boundary should satisfy $\varrho/\varsigma \geq 0$. This condition sets the bound $\mathrm{s}\geq{\rm n}\left(3(1-\nu)/({1+\nu})+2{\rm n} (1-2\nu)/(1+\nu)\right)^{-1}$. As expected, decreasing the shear index is analogous to decreasing the effective elasticity on the body, and thus the compactness decreases with respect to the ${\rm s = n}$ case.

\section{Discussion and Outlook}
Our model of elastic star provides the first example of a self-gravitating material object with scale invariance within GR. 
While the radial stability of these solutions is already promising, future work should assess their full linear stability beyond spherical symmetry.

An interesting avenue would be to link our framework to an underlying microscopic description of elastic matter at (ultra)nuclear densities, as those expected in neutron star cores and during gravitational collapse. Indeed, solid phases of matter may be relevant for astrophysical compact objects: while degenerate fermions behave as a weakly-interacting gas at relatively small densities, nuclear interactions and QCD effects become crucial inside relativistic stars, and the perfect fluid idealization eventually breaks down.
This is certainly the situation in the crust of a neutron star~\cite{Chamel:2008ca,Suleiman:2021hre}, while the fundamental constituents in the core are still largely unknown~\cite{Lattimer:2004pg}.

Our formalism can be naturally extended in two important directions. Firstly,
an extension beyond spherical symmetry is underway. This would allow us to study rotating solutions, in particular observable quantities such as their moment of inertia and spin-induced quadrupole moment. It would be interesting to assess whether elastic stars feature approximately universal (i.e., EoS independent) relations among these quantities, as in the case of perfect fluid neutron stars~\cite{Yagi:2016bkt}, since these relations have important astrophysical consequences.
Furthermore, an extension beyond spherical symmetry would allow building generically deformed elastic objects, which are not allowed in the fluid limit~\cite{Raposo:2020yjy}.
Overall, we anticipate that it should be possible to study the entire multipolar structure of these solutions, which is richer than fluid stars and also different from BHs, providing more realistic models and a way to discriminate compact elastic stars from BHs or perfect fluid neutron stars~\cite{Raposo:2018rjn,Cardoso:2019rvt,Maggio:2021ans}.

The second important extension concerns the nonlinear time evolution for elastic objects. This is a major obstacle for various models of BH mimickers that are either phenomenological or do not have a well-posed time evolution~\cite{Cardoso:2019rvt}.
Remarkably, the initial-value problem in spherical symmetry in our model can be shown to be strictly hyperbolic, at least in Minkowski spacetime~\cite{Alho:2023ris}.
This provides a promising starting point to study the full nonlinear evolution of elastic stars within our framework.
An outstanding question concerns the merger of two elastic stars in this model: due to the absence of a maximum mass, it is possible that the merger remnant is always a (stable) heavier and larger star living in the same linear mass-radius diagram, thus preventing BH formation.
Finally, a healthy time evolution beyond spherical symmetry would allow studying the tidal perturbations of elastic stars and their ringdown, both of which have direct consequences for gravitational-wave astronomy.
Based on our results, we expect the phenomenology of tidal perturbations and the ringdown of physically viable elastic stars to be qualitatively more similar to that of perfect fluid neutron stars than to BHs, but it would anyway quantitatively depart from the standard perfect fluid case.
Thus, it would be interesting to study the inspiral, merger, and post-merger phase of elastic star coalescences in this model, to distinguish the emitted gravitational-wave signal from that of binary BHs or perfect fluid neutron stars.
We hope to report on these followups elsewhere.

\acknowledgments
   A.A.\ and J.N.\ were partially
supported by FCT/Portugal through CAMGSD, IST-ID, projects UIDB/04459/2020 and UIDP/04459/2020, and also by the H2020-MSCA-2022-SE project EinsteinWaves, GA no.~101131233.
P.P.\ acknowledges financial support provided under the European Union's H2020 ERC, Starting 
Grant agreement no.~DarkGRA--757480, and under the MIUR PRIN and FARE programmes (GW-NEXT, CUP:~B84I20000100001), and support from the Amaldi Research Center funded by the MIUR program `Dipartimento di Eccellenza" (CUP:~B81I18001170001).
G.R.\ was supported by the Center for Research and Development in Mathematics
and Applications (CIDMA) through the Portuguese Foundation for Science and Technology (FCT - Fundação para a Ciência e a Tecnologia), references UIDB/04106/2020 (\href{https://doi.org/10.54499/UIDB/04106/2020}{https://doi.org/10.54499/UIDB/04106/2020}) and UIDP/04106/2020 (\href{https://doi.org/10.54499/UIDP/04106/2020}{https://doi.org/10.54499/UIDP/04106/2020}) and by the 5th Individual CEEC program, reference 10.54499/2022.04182.CEECIND/CP1720/CT0004 (\href{https://doi.org/10.54499/2022.04182.CEECIND/CP1720/CT0004}{https://doi.org/10.54499/2022.04182.CEECIND/CP1720/CT0004}).  
 
%
\appendix
\section{Derivation of a general scale-invariant EoS}\label{app:EoSgeneral}

In this appendix, we show that Eq.~\eqref{eq:scaleinvEOS}, obtained from the straightforward generalization of the fluid linear EoS~\eqref{rhoISO}, is in fact the most general form for a scale-invariant EoS. 

As discussed in the main text, for a scale-invariant matter source in spherical symmetry, the energy density, the radial and the tangential pressures transform as
Eq.~\eqref{rescal}, and the Tolman-Oppenheimer-Volkoff~(TOV) system of equations is invariant under the scaling transformation
\begin{equation}\label{scaling_r}
    r\to A^{-1}\tilde{r}\,.
\end{equation}
%
%
%

In the perfect fluid case, scale invariance is achieved for a subclass of models~\cite{TC1971} with constant adiabatic index $\gamma=d\ln{\widehat{p}_\mathrm{iso}}/d\ln{\varrho}$, and a linear EoS, see Eq.~\eqref{EoSFluid}.
This is in turn equivalent to $\widehat{\rho}(\varrho)$ being a positively homogeneous function of degree $\gamma$, and the TOV system~\eqref{TOVeq} is invariant under the scaling transformation~\eqref{scaling_r} and
\begin{equation}\label{scaling_varrho}
\varrho(r) \to A^{2/\gamma} \tilde{\varrho}(\tilde{r}).
\end{equation}

Extending to the elastic setting, the fact that $\rho$, $p_{\rm rad}$, and $p_{\rm tan}$ scale in the same fashion [see Eq.~\eqref{rescal}] suggests that we look for an EoS in which $\rho$ is a \emph{linear} function of $p_{\rm rad}$, and $p_{\rm tan}$,
\begin{equation}
\widehat{\rho}(\varrho,\varsigma)  = \frac{\rm n}3 \left( a\,\widehat{p}_{\rm rad}(\varrho,\varsigma) + b\,\widehat{p}_{\rm tan} (\varrho,\varsigma)\right)   \,,\label{linear} 
\end{equation}
where $n$ and $\gamma$ are related by $\gamma = 1+1/n$, and the constants $a,b$ are constrained by $a+b=3$ in order to recover the fluid limit, Eq.~\eqref{EoSFluid}. Substituting the definitions of the pressures [Eqs.~\eqref{eq:EoS}], the above equation can be written as a linear PDE for $\widehat{\rho}(\varrho,\varsigma)$, whose solution is
\begin{equation}\label{rho_scainv}
    \widehat{\rho}(\varrho,\varsigma)= \mathrm{n}\mathcal{K}\varsigma^{1+\frac1{\rm n}} h(y) \,,
\end{equation}
where $y=\varrho^{(3-a)/2}/\varsigma$ and $h(y)$ is a free function.
From the definition of $\varsigma$ [Eq.~\eqref{ODEvarsigma}], it follows from the scaling transformations~\eqref{scaling_r},~\eqref{scaling_varrho} that
\begin{equation}
    \varsigma(r)\to A^{2/\gamma}\tilde{\varsigma}(\tilde{r})\,.
\end{equation}
That is, $\varrho$ and $\varsigma$ transform in the same fashion under the scaling transformation. Thus, Eq.~\eqref{rescal} imposes that $h(y)$ must be invariant under the scaling, and hence $a=1$, $b=2$ is the only choice, leading to the ansatz~\eqref{rhoISO} in the main text, i.e. 
\begin{equation}
\widehat{\rho}(\varrho,\varsigma) = {\rm n} \widehat{p}_{\rm iso}(\varrho,\varsigma) = \frac{\rm n}3 \left( \widehat{p}_{\rm rad}(\varrho,\varsigma) + 2\widehat{p}_{\rm tan} (\varrho,\varsigma)\right)\,,
\end{equation}
and the scale-invariant quantity $y(r)\equiv\varrho(r)/\varsigma(r)$ is the spherically symmetric shear variable. Just as in the perfect fluid case, the above choice is equivalent to $\widehat{\rho}(\varrho,\varsigma)$ being a \emph{positively homogeneous function of degree $\gamma$},    
\begin{equation}\label{f_function}
\widehat{\rho}(\varrho,\varsigma) = \varsigma^{1+\frac1{\rm n}} \widehat{\rho}(\varrho/\varsigma,1)
= \mathrm{n}\mathcal{K}\varsigma^{1+\frac1{\rm n}} h(y) \,,
\end{equation}
which justifies Eq.~\eqref{eq:scaleinvEOS} used in the main text.
\bibliographystyle{utphys}
\bibliography{biblio}

\end{document}